\documentclass[english]{article}
\usepackage[T1]{fontenc}
\usepackage[latin9]{inputenc}
\usepackage{textcomp}
\usepackage{amsmath}
\usepackage{amssymb}
\usepackage{esint}
\usepackage{babel}
\begin{document}
\title{Adiabatic driving and geometric phases in classical systems }
\author{A. D. Berm\'udez Manjarres\thanks{ad.bermudez168@uniandes.edu.co}}
\maketitle
\begin{abstract}
We study the concepts of adiabatic driving and geometric phases of
classical integrable systems under the Koopman-von Neumann formalism.
{\normalsize{}In close relation to what happens to a quantum state,
a classical }Koopman-von Neumann{\normalsize{} eigenstate will acquire
a geometric phase factor $exp\left\{ i\Phi\right\} $ after a closed
variation of the parameters $\lambda$ in its associated Hamiltonian.
The explicit form of $\Phi$ is then derived for integrable systems,
and its relation with the Hannay angles is shown. Additionally, we
use quantum formulas to write a classical adiabatic gauge potential
that generates adiabatic unitary flow between classical eigenstates,
and we explicitly show the relationship between the potential and the
classical geometric phase.}{\normalsize\par}
\end{abstract}

\section{Introduction}

In quantum mechanics, the Berry phase is the change in the phase of
an eigenstate throughout a cyclic and adiabatic path in the parameter
space of the Hamiltonian \cite{berry}. This change has nothing to
do with dynamics, its nature is purely geometric. Several generalizations
have been given for the Berry phase. We mention the non-Abelian treatment
given by Wilczeck and Zee for the case of Hamiltonian with degenerate
spectrum\cite{WZ}. 

There is an analogous holonomic effect for integrable systems in classical
mechanics. This effect is called the Hannay angles, and it is the
change in the angle variables when the parameters in the classical
Hamiltonian makes an adiabatic closed circuit in parameter space \cite{hannay,hannay2,hannay3}.
It is known that, for integrable systems, the Hannay angles are the
classical limit of the\textbf{ }Berry phases \cite{hannay2}. 

However, despite their similarities, these two holonomy effects are
usually treated using different mathematical formalisms. A Hilbert
space formalism for the Berry phases and phase space functions for
the Hannay angles (though there are some notable exceptions \cite{berryhannay,berryhannay2}).

Here we will study the classical geometric phases (and adiabatic driving,
see below) in the same mathematical language of quantum mechanics.
The above will be possible due to the formulation of classical mechanics
known as the Koopman-von Neumann Theory (KvN) \cite{KvN1,KvN2,KvN3,KvN4,KvN5}.
The KvN theory is an operational version of classical mechanics akin
to quantum mechanics. It is composed of a Hilbert space, a Schr\"{o}dinger-like
equation of motion, and other features that are reminiscent of quantum
mechanics.

Not only is the KvN Theory interesting in itself in a purely classical
context, but it is also connected to the quantum-classical correspondence.
The KvN theory is related to geometric quantization \cite{klein}
and the classical limit of the Wigner function \cite{wignerKvN,wignerKvN2}. 

To the author's knowledge, the first to consider the geometric phases
of KvN eigenfunctions was de Polavieja \cite{polavieja,polavieja 2}.
However, reference \cite{polavieja} seems to be not well known, and
reference \cite{polavieja 2} is not so explicit about the issue.
Hence, I believe it is warranted to give a more detailed exposition
of the topic.

The KvN formalism will also allow us to give a treatment of the adiabatic
driving of classical (integrable) states in a parallel fashion to
quantum mechanics. The central object we will consider is the so-called
adiabatic gauge potential \cite{adriving,adriving2,adriving3,adriving4,adriving5,adriving6}. 

In the next section, the necessary elements of KvN formalism will be
exhibited. It will be shown how to associate a Hilbert space to the
classical phase space, the inner product in this Hilbert space, and
the equation of motion of the state vector (or wave functions). Of
special interest will be the form of the KvN waves when angle-action
variables are used \cite{angleaction,angleaction2,prigogine}. It
will be shown, using formulas from quantum mechanics, that the geometric
phase factor acquired by the KvN states is related to the Hannay angle.

In section 3 we use quantum formulas to define the adiabatic gauge
potential in the KvN theory. The potential generates a unitary flow
of the KvN states which correspond to a canonical flow in the usual
language of Hamiltonian mechanics. The potential will be related to
generating functions of the Lie-Deprit perturbation theory. We show
the connection between the Yang-Mills curvature of the potential and
the Hannay curvature.

\section{The Koopman-von Neumann Formalism}

The starting point of KvN theory is Liouville's equation for the
evolution of probability density in phase space, 

\begin{equation}
\frac{\partial\rho(p,q,t)}{\partial t}=-\{\rho,H(p,q)\},\label{eq:1}
\end{equation}
where $p=(p_{1},....,p_{n})$, $q=(q_{1},...,q_{n})$, $H(p,q)$ is
the Hamiltonian of the system, and $\{,\}$ is the Poisson bracket
\cite{arnold}. From this, we can write classical (statistical) mechanics
in the same mathematical language of quantum mechanics. This is, we
will have a set of (classical) state vectors or waves functions and
the inner product between those states (i,e. A Hilbert space), a ``Hamiltonian''-like
operator, and a Schr\"{o}dinger-like equation of motion. The procedure
is as follows: first, the KvN wavefunctions are defined by the relation 

\begin{equation}
\psi^{*}(p,q,t)\psi(p,q,t)=\rho(p,q,t),\label{eq:2}
\end{equation}
Inserting equation (\ref{eq:2}) into equation (\ref{eq:1}) it can
be shown that the KvN wavefunctions satisfy the Schr\"{o}dinger-like equation
\cite{KvN3}

\begin{eqnarray}
i\frac{\partial\psi}{\partial t} & = & \hat{L}\psi,
\end{eqnarray}
where the role of the ``Hamiltonian'' operator is played by the
Liouvillian defined by 
\begin{equation}
\hat{L}=-i\{,H(p,q)\}.
\end{equation}
The set of all square-integrable functions of phase space defines the
Hilbert space $\mathcal{H}_{c}$ of the classical wavefunctions. The
inner product in $\mathcal{H}_{c}$ is defined by

\begin{equation}
\left\langle \varphi,\psi\right\rangle =\int\varphi^{*}\psi\,d^{n}q\,d^{n}p,
\end{equation}
where the integral is taken over all phase space. It can be shown
that $\hat{L}$ is Hermitian under this inner product. 

Suppose the Liouvillian depends on a set of external parameters $\hat{L}(\lambda),$
where $\lambda=(\lambda_{1},\lambda_{2},...,\lambda_{N})$, and consider
the eigenvalue equation 
\begin{equation}
\hat{L}(\lambda)\psi_{n}(\lambda)=l_{n}(\lambda)\psi_{n}(\lambda).
\end{equation}
If the parameters vary slowly enough so the conditions of the adiabatic
theorem are met, then the state of the system will remain as an eigenfunction
of the instantaneous Liouvillian. From the work of Berry, we can expect
that after a closed loop in parameter space the original eigenfunctions
acquires a geometric phase
\begin{equation}
\psi_{n}\longrightarrow phase\;factor\:\psi_{n}.
\end{equation}
The spectrum of the Liouvillian is degenerate and the eigenspaces
are of infinite dimension, as we will see. Thus, we will use the Wilzeck-Zee
formula to calculate this phase factor.

\subsection{Integrable systems}

We are interested in integrable Hamiltonians in the sense of Liouville-Arnold
where the system can be described in terms of action-angle variables
$(\phi,I)$ and the Hamiltonian depends only on the actions, i.e.,
$H=H(I)$ \cite{arnold}. We restrict ourselves to one degree of freedom,
but our results can be generalized without a problem. In this case,
Liouvillian reduces to the simplified form

\begin{equation}
\hat{L}=-i\omega\frac{\partial}{\partial\phi},
\end{equation}
where the frequency is given by

\begin{equation}
\omega=\frac{\partial H(I)}{\partial I}.
\end{equation}
The eigenfunctions $\psi_{n}(I,\phi)$ of the Liouvillian 

\begin{equation}
\hat{L}\psi_{n}=l_{n}\psi_{n},\label{eq:L9}
\end{equation}
are given by \cite{angleaction2}

\begin{equation}
\psi_{n}=\frac{1}{\sqrt{2\pi}}\delta(I-I\text{\textasciiacute)}\,e^{in\phi},\label{eq:W}
\end{equation}
where $n\in\mathbb{Z},$ and the eigenvalues are

\begin{equation}
l_{n}=n\omega.
\end{equation}

We can see that the eigenfunctions and eigenvalues depend on the discrete
index $n$, but the eigenfunctions also depend on the continuous $I$.
Hence, the spectrum of the Liouvillian is uncountably degenerated.
To avoid dealing with Dirac's deltas, we will discretize the variable
$I$ by defining the following functions

\begin{equation}
\psi_{k,n}(I,\phi)=\frac{1}{\sqrt{2\pi}}f(I)_{k}\,e^{in\phi},\label{eq:phif}
\end{equation}
where

\begin{eqnarray}
f(I)_{k} & = & \begin{cases}
0 & if\;I<k,\\
f(I) & if\;k<I<k+\delta k,\\
0 & if\;k+\delta k<I,
\end{cases}\label{eq:f3}\\
1 & = & \int_{0}^{\infty}dI\:f(I)_{k}^{2}.\label{eq:3}
\end{eqnarray}

The exact form of $f(I)$ will not be important. The $\psi_{k,n}(I,\phi)$
are eigenfunctions of the Liouvillian, and we will use them to compute
the geometric phase in the next section. The discretizing procedure
described above is equivalent to Weyl's eigendifferentials \cite{messiah},
a concept that has already been used to define the geometric phase
for quantum systems with continuous spectrum \cite{continuous}. 

From the definition (\ref{eq:3}) and the non-intersecting nature
of the Liouville-Arnold Tori, it follows that the functions $f(I)_{k}$
obey the orthogonality relation

\begin{equation}
\int_{0}^{\infty}dI\:f(I)_{k}f(I)_{k\text{\textasciiacute}}=\delta_{kk\text{\textasciiacute}}.
\end{equation}

Finally, notice that we can write the 

\begin{equation}
\int dq\,dp=\int_{0}^{\infty}dI\int_{0}^{2\pi}d\phi.
\end{equation}

\subsection{Geometric phases}

In this section, we will calculate the holonomy associated with an
adiabatic transport of the KvN eigenfunctions of an integrable system
through a closed loop in parameter space \cite{polavieja,polavieja 2}.
For this, we will use Wilzeck-Zee formulas for the geometric phase
of degenerate systems \cite{WZ}. 

Let $H(I(\lambda),\lambda)$ be a classical Integrable hamiltonian
with dependency on some external parameters $\lambda=(\lambda_{1},\lambda_{2},...,\lambda_{n})\in\mathit{\mathcal{M}}$,
where \textit{$\mathcal{M}$ }a smooth manifold, and we assume the
variation in parameter space obeys the conditions of the adiabatic
quantum theorem. For our KvN waves, the matrix components of the Wilzeck-Zee
non-abelian potential for the nth eigenvalue subspace are given by

\begin{eqnarray}
A_{kk\text{\textasciiacute}}^{(n)} & = & i\left\langle \psi_{k,n},d\psi_{k',n}\right\rangle ,\label{eq:wz}
\end{eqnarray}
were $d_{\lambda}$ stands for the exterior derivative in parameter
space. Once computed this 1-form, we can calculate the holonomy using
a path-ordered exponential

\begin{equation}
\psi_{n}(\lambda(t))=Pe^{\oint_{\lambda}\tilde{A}^{(n)}}\psi_{k,n}(\lambda(t_{0})).
\end{equation}

Usually, the computation of Path-ordered exponentials is not an easy
task. However, for the case at hand, we will see that this computation
is greatly simplified. 

Using the wave function (\ref{eq:W}), we have that

\begin{eqnarray}
A_{kk\text{\textasciiacute}}^{(n)} & = & \frac{i}{2\pi}\int_{0}^{\infty}dI\int_{0}^{2\pi}d\phi\:f(I)_{k}e^{-in\phi_{\lambda}}d_{\lambda}(f(I)_{k\text{\textasciiacute}}e^{in\phi})\nonumber \\
 & = & \frac{i}{2\pi}\int_{0}^{\infty}dIf(I)_{k}f\text{\textasciiacute}(I)_{k\text{\textasciiacute}}\int_{0}^{2\pi}d\phi\:d_{\lambda}I\nonumber \\
 &  & -\frac{1}{2\pi}\int_{0}^{\infty}dI\int_{0}^{2\pi}d\phi\:f(I)_{k}f(I)_{k\text{\textasciiacute}}d_{\lambda}\phi.
\end{eqnarray}
The first term in the last line vanishes identically due to Liouville's
theorem \cite{arnold}, namely

\begin{equation}
\left\langle d_{\lambda}I(\lambda)\right\rangle =0,
\end{equation}
where the torus average is defined by

\begin{align}
\left\langle \cdots\right\rangle  & =\frac{1}{2\pi}\int_{0}^{\infty}(\cdots)\,d\phi\nonumber \\
 & =\int\,(\cdots)\,\delta(I(p,q,\lambda)-I_{0})\frac{dp\,dq}{2\pi}.
\end{align}
The components of the Wilzeck-Zee potential are then given by the
reduced expression

\begin{eqnarray}
A_{kk\text{\textasciiacute}}^{(n)}(\lambda) & = & -n\left(\int_{0}^{\infty}dI\;f(I)_{k}f(I)_{k\text{\textasciiacute}}\right)\left(\frac{1}{2\pi}\int_{0}^{2\pi}d\phi\:d_{\lambda}\phi\right)\nonumber \\
 & = & -n\delta_{kk\text{\textasciiacute}}\left\langle d_{\lambda}\phi(\lambda)\right\rangle .\label{eq:A}
\end{eqnarray}
The diagonal nature of the (\ref{eq:A}) shows that there is no internal rotation in the eigenspace, each wavefunction $\psi_{n}$ remains in
the torus with the same value of $I$ as it started, as expected from
the classical adiabatic theorem \cite{arnold}. So, after a complete circuit in parameter space, we can directly write 

\begin{equation}
\psi_{n}(\lambda(t))=e^{in\Phi}\psi_{n}(\lambda(t_{0})),
\end{equation}
with

\begin{alignat}{1}
\Phi & =-\Delta\phi_{Hannay}=-\oint_{C}\left\langle d_{\lambda}\phi\right\rangle .
\end{alignat}
We see that the geometric phase factor acquired by the KvN wave is
proportional to the Hannay angle \cite{hannay}.

\section{Adiabatic driving and the non-Abelian adiabatic gauge potential}

For this section, it will be convenient to consider
the action $I$ in the Liouvillian as a parameter. In that case, the
eigenfunctions of $\hat{L}$ are

\begin{equation}
\psi_{n}=\frac{1}{\sqrt{2\pi}}e^{in\phi}.
\end{equation}
The functions $\psi_{n}(\phi)$ do not form a complete set in phase
space, but we can still represent any function $f(I,\phi)$ as a superposition
of the $\psi_{n}(\phi)$ 

\begin{equation}
f(I,\phi)=\frac{1}{\sqrt{2\pi}}\sum_{n}a_{n}(I)e^{in\phi},
\end{equation}
where

\[
a_{n}(I)=\frac{1}{\sqrt{2\pi}}\int_{0}^{2\pi}d\phi\,f(I,\phi)e^{-in\phi}.
\]
This corresponds to the usual Fourier expansion of $f(I,\phi)$ \cite{prigogine}.
With this choice, the inner product becomes

\[
\left\langle \phi,\varphi\right\rangle =\int_{0}^{2\pi}\phi^{*}\varphi\,d\phi.
\]

Now, we have the following situation in quantum mechanics: if the
initial state of the system if an eigenvector of the Hamiltonian $\hat{H}(\lambda)\left|\psi_{n}(\lambda)\right\rangle =E(\lambda)\left|\psi_{n}(\lambda)\right\rangle $,
and we make a finite adiabatic change $\lambda\rightarrow\lambda'$
so the instantaneous state obeys the equation $\hat{H}(\lambda')\left|\psi_{n}(\lambda')\right\rangle =E(\lambda')\left|\psi_{n}(\lambda')\right\rangle $,
then the initial and final states are related by a unitary transformation
of the form \cite{adriving} 
\begin{equation}
\left|\psi_{n}(\lambda')\right\rangle =P\exp\left(i\int_{\lambda}^{\lambda'}\hat{\mathcal{A}}\right)\left|\psi_{n}(\lambda)\right\rangle ,
\end{equation}
where the operator-valued non-Abelian connection 1-form $\mathcal{A}$
is known as the adiabatic gauge potential. 

In our KvN notation, the components of the adiabatic potential are
defined by the equations

\begin{eqnarray}
\left\langle \psi_{n},\hat{\mathcal{A}}\psi_{n}\right\rangle  & = & 0,\label{berry-simon}\\
\left\langle \psi_{m},\hat{\mathcal{A}}\psi_{n}\right\rangle  & = & -i\frac{\left\langle \psi_{m},d_{\lambda}\hat{L}\psi_{n}\right\rangle }{l_{n}-l_{m}},\label{rotation}
\end{eqnarray}
where $d_{\lambda}\hat{L}=-i\left\{ \cdot,d_{\lambda}H(\lambda)\right\} $,
and the $l_{n}=l_{n}(\lambda)$ are the instantaneous eigenvalues
of the Liouvillian. The action of $\hat{\mathcal{A}}$ on the eigenfunctions
is

\begin{equation}
\hat{\mathcal{A}}e^{in\phi}=-id_{\lambda}e^{in\phi}=-i\sum_{m\neq n}\frac{\left\langle \psi_{m},d_{\lambda}\hat{L}\psi_{n}\right\rangle }{\omega(n-m)}e^{im\phi},\label{defA}
\end{equation}
where the frequency depends on $\lambda$. The equation (\ref{rotation})
gives the rotation of the eigenfunctions after a small change in the
parameters, and (\ref{berry-simon}) is the Berry-Simon condition
for parallel transport \cite{simon,cruchinski}. The potential 1-form
has several integral representations, we will use the following one
that arises from the van Vleck-Primas perturbation theory \cite{adriving}

\begin{align}
\hat{\mathcal{A}} & =-\lim_{T\rightarrow\infty}\frac{1}{T}\int_{0}^{T}dt\int_{0}^{t}ds\,e^{-is\hat{L}}(d_{\lambda}\hat{L})e^{is\hat{L}}.\label{integralA}
\end{align}

Now, we want to write the components of the potential as

\begin{equation}
\hat{\mathcal{A}}(\lambda)=-i\left\{ ,W(I,\theta,\lambda)\right\} ,
\end{equation}
where \{,\} is the Poisson bracket, and the generating function $W_{\mu}$
is to be determined. We can compute (\ref{defA}) directly to obtain

\begin{eqnarray}
\hat{\mathcal{A}}(\lambda)e^{in\phi} & = & i\frac{n}{\omega}\frac{\partial}{\partial I}\sum_{m\neq n}\left[\frac{1}{2\pi(n-m)}\int_{0}^{2\pi}d\phi\:e^{i(n-m)\phi}d_{\lambda}H\right]e^{im\phi}.\label{A1}
\end{eqnarray}
To continue we now make the change $m\rightarrow m+n$ in the summation
index, to get

\begin{eqnarray}
\hat{\mathcal{A}}(\lambda)e^{in\phi} & = & -i\frac{n}{\omega}e^{im\phi}\frac{\partial}{\partial I}\sum_{m\neq0}\left[\frac{1}{2\pi m}\int_{0}^{2\pi}d\phi\:e^{-im\phi}d_{\lambda}H\right]e^{im\phi}.\label{A2}
\end{eqnarray}
The term $\sum_{m\neq0}\left[\frac{1}{2\pi m}\int_{0}^{2\pi}d\phi\:e^{-im\phi}d_{\lambda}H\right]e^{im\phi}$
is the the Fourier expansion of $\int d_{\lambda}H\,d\phi$ but without
the $m=0$ coefficient, the secular term of $d_{\lambda}H$. This
means that the actual function expanded is $\int(d_{\lambda}H-\left\langle d_{\lambda}H\right\rangle )\,d\phi$.
Therefore, we can continue as

\begin{eqnarray}
\hat{\mathcal{A}}(\lambda)e^{in\phi} & = & n\frac{1}{\omega}e^{in\phi}\frac{\partial}{\partial I}\left[\int d\phi\:\left(d_{\lambda}H-\left\langle d_{\lambda}H\right\rangle \right)\right]\nonumber \\
 & = & -i\left\{ e^{in\phi},\frac{1}{\omega}\int d\phi\:\left(\left\langle d_{\lambda}H\right\rangle -d_{\lambda}H\right)\right\} .
\end{eqnarray}
Hence, we can identify the generating function as

\begin{equation}
W(I,\theta,\lambda)=\frac{1}{\omega}\int d\phi\:\left(\left\langle d_{\lambda}H\right\rangle -d_{\lambda}H\right).\label{W}
\end{equation}
which corresponds to the first-order generating function in the Lie
series perturbation theory of Hamiltonian mechanics \cite{ferraz}.
Notice that $W$ has to be written in terms of the original set of
angle action variables. The above exemplifies the close relationship
between the van Vleck-Primas and the Lie-Deprit perturbation theories
\cite{hamiltonianp}.

Alternatively, we can use Eq (\ref{integralA}) to compute $\hat{\mathcal{A}}$
as follows:

\begin{align}
\left\langle \psi_{m},\hat{\mathcal{A}}\psi_{n}\right\rangle  & =-\lim_{T\rightarrow\infty}\frac{1}{2\pi T}\int_{0}^{T}dt\int_{0}^{t}ds\,\int_{0}^{2\pi}d\phi\,e^{-im\phi}\left\{ e^{in\phi},d_{\lambda}H\right\} e^{is(n-m)\omega}\nonumber \\
 & =i\lim_{T\rightarrow\infty}\frac{1}{T}\frac{\partial}{\partial I}\int_{0}^{T}dt\frac{n}{(n-m)\omega}e^{i(n-m)\phi}d_{\lambda}H.\label{A}
\end{align}
In the last line of Eq (\ref{A}), we can change from the time average
$\lim_{T\rightarrow\infty}\frac{1}{T}\int_{0}^{T}dT$ to a torus average
$\frac{1}{2\pi}\int_{0}^{2\pi}d\phi$ \cite{arnold 2}, so the expression
is identical to (\ref{A1}).

For closed paths, the action of the adiabatic potential is of the
form 

\emph{
\begin{equation}
Pe^{i\oint_{\gamma}\hat{\mathcal{A}}}\psi_{n}=e^{i\Phi_{n}}\psi_{n},\label{lemma}
\end{equation}
}where $\Phi_{n}$ is the geometric phase acquired by the KvN wave.
The information about the geometric phase is encoded in the Yang-Mills
curvature associated with $\hat{\mathcal{A}}$. The components of
curvature are given by 

\begin{equation}
\hat{F}_{\mu\nu}(\lambda)=\partial_{\mu}\hat{\mathcal{A}}_{\upsilon}-\partial_{\upsilon}\hat{\mathcal{A}}_{\mu}-i\left[\hat{\mathcal{A}}_{\mu},\hat{\mathcal{A}}_{\upsilon}\right].\label{curvature}
\end{equation}
The curvature 2-form is diagonal in the original basis of eigenfunctions
and its integral gives the geometric phase \cite{adriving}

\begin{equation}
\int\left\langle \psi_{m},\hat{F}\psi_{n}\right\rangle =\delta_{nm}\,nth\,Berry\,phase.
\end{equation}
As we saw in the previous section, the geometric phase of the KvN
theory is related to the classical Hannay angles. The holonomy of
$\hat{\mathcal{A}}$ agrees with this, as we will see.

We now compute directly the non-vanishing matrix elements of the curvature
using its definition (\ref{curvature}). The term $\left\langle \psi_{m},\partial_{\mu}\hat{\mathcal{A}}_{\upsilon}\psi_{n}\right\rangle $
gives

\begin{eqnarray*}
\left\langle \psi_{n},\partial_{\mu}\mathcal{A}_{\nu}\psi_{n}\right\rangle  & = & -i\frac{1}{2\pi}\int_{0}^{2\pi}d\phi\:e^{-in\phi}\left\{ e^{in\phi},\partial_{\mu}W_{\nu}\right\} \\
 & = & \frac{1}{2\pi}\int_{0}^{2\pi}d\phi\,\frac{\partial(\partial_{\mu}W_{\nu})}{\partial I}=\left\langle \frac{\partial(\partial_{\mu}W_{\nu})}{\partial I}\right\rangle =0,
\end{eqnarray*}
where the last line follows from the fact that $W_{\nu}$ has no secular
term (see Eq (\ref{W})). $\left\langle \psi_{n},\partial_{\nu}\hat{\mathcal{A}}_{\mu}\psi_{n}\right\rangle $
vanishes for the same reason. We can proceed with the commutator as 

\begin{eqnarray}
-i\left\langle \psi_{n},\left[\hat{\mathcal{A}}_{\mu},\hat{\mathcal{A}}_{\upsilon}\right]\psi_{n}\right\rangle  & = & -i\frac{1}{2\pi}\int_{0}^{2\pi}d\phi\,e^{-in\phi}\left\{ e^{in\phi},\left\{ W_{\nu},W_{\mu}\right\} \right\} We\nonumber \\
 & = & \frac{\partial}{\partial I}\left\langle \left\{ W_{\nu},W_{\mu}\right\} \right\rangle .\label{ws}
\end{eqnarray}
We show in the appendix that (\ref{ws}) gives the Hannay curvature.

\subsection{Example: Generalized oscillator}

The Hamiltonian of the generalized oscillator is 
\begin{equation}
H=\frac{1}{2}\left(Xq^{2}+2Yqp+Zp^{2}\right),
\end{equation}
where $X,Y,$ and $Z$ are the parameters to be varied. A possible
canonical transformation to angle\textendash action variables is \cite{cruchinski}

\begin{align}
q & =\sqrt{\frac{2IZ}{\omega}}\cos\phi,\nonumber \\
p & =-\sqrt{\frac{2IZ}{\omega}}(\frac{Y}{Z}\cos\phi+\frac{\omega}{Z}\sin\phi),\label{CT}
\end{align}
where the frequency is $\omega=\sqrt{(XZ-Y^{2})}$. The variation
of the Hamiltonian can be written in terms of the original angle\textendash action
as follows:
\begin{align}
\frac{\partial H}{\partial Y} & =pq=-(\frac{2ZI}{\omega})(\frac{Y}{Z}\cos^{2}\phi+\frac{\omega}{Z}\sin\phi\cos\phi),\nonumber \\
\frac{\partial H}{\partial X} & =\frac{ZI}{\omega}\cos^{2}\phi=\frac{ZI}{2\omega}(1+\cos(2\phi)),\nonumber \\
\frac{\partial H}{\partial Z} & =\frac{ZI}{\omega}(\left[\tfrac{Y}{Z}\right]^{2}\cos^{2}\phi+\left[\tfrac{\omega}{Z}\right]^{2}\sin^{2}\phi+\tfrac{Y\omega}{Z^{2}}\sin2\phi).\label{dh}
\end{align}
Using (\ref{dh}) into (\ref{W}), we get the generating functions

\begin{align}
W_{Y} & =\frac{ZI}{2\omega^{2}}(\frac{Y}{Z}\sin2\phi-\frac{\omega}{Z}\cos2\phi),\nonumber \\
W_{X} & =-\frac{ZI}{4\omega^{2}}\sin(2\phi),\nonumber \\
W_{Z} & =-\frac{ZI}{4\omega^{2}}(\left[\tfrac{Y}{Z}\right]^{2}\sin2\phi-\left[\tfrac{\omega}{Z}\right]^{2}\sin2\phi-2\tfrac{%
Y\omega}{Z^{2}}\cos2\phi).\label{Ws}
\end{align}
We need the following Poisson brackets to compute the components of
the curvature,

\begin{align}
\{W_{Y},W_{X}\} & =\frac{Z}{8\omega^{3}}\left\{ I\sin(2\phi),I\cos2\phi\right\} =-\frac{ZI}{4\omega^{3}},\nonumber \\
\{W_{X},W_{Z}\} & =-\frac{YI}{4\omega^{3}},\nonumber \\
\{W_{Z},W_{Y}\} & =-\frac{XI}{4\omega^{3}}.
\end{align}
Collecting everything, we obtain the diagonal matrix elements of the
curvature 2-form

\begin{equation}
\left\langle \psi_{n},\hat{F}\psi_{n}\right\rangle =\frac{n}{4\omega^{3}}\left(XdY\wedge dZ+YdZ\wedge dX+ZdX\wedge dY\right).\label{curvature-1}
\end{equation}
The expression (\ref{curvature-1}) agrees with the Hannay curvature
of the system \cite{hannay}.

\section{Conclusion}

We gave a treatment of the classical geometric phases and the adiabatic
gauge potential in the context of the Koppman-von Neumann formulation
of classical mechanics. In both cases, using quantum formulas leads
to well-established expressions from Hamiltonian mechanics.

Our treatment was restricted to systems with one degree of freedom,
but it can be easily generalized to higher dimensions for completely
integrable Hamiltonians. 

There is, in principle, no reason to restrict the analysis of geometric
phases and adiabatic driving of KvN states to integrable systems.
After all, all the relevant formulas come from quantum mechanics and
they are independent of the form of the Hamiltonian/Liouvillian. A
natural research line is then to give an analogous analysis for classically
chaotic systems and to compare it with previous results \cite{robbins,jar}.

\subsection*{Appendix}

In this appendix, we will show that
\begin{align}
\int\left\langle \psi_{n},\hat{F}\psi_{n}\right\rangle  & =\frac{\partial}{\partial I}\int\left\langle \left\{ W_{\nu},W_{\mu}\right\} \right\rangle d\lambda_{\mu}\wedge d\lambda_{\nu}\nonumber \\
 & =-\frac{\partial}{\partial I}\int\left\langle d_{\lambda}\phi(\lambda)\wedge d_{\lambda}I(\lambda)\right\rangle =-\Delta\phi_{Hannay}.\label{dI}
\end{align}

The proof is as follows: Let $H(\lambda)$ be a family of integrable
Hamiltonians and suppose we know how to write the angle-action variables
$(I(\lambda),\phi(\lambda))$. We can relate the action-angle variables
of two infinitesimally close Hamiltonians $H(\lambda)$ and $H(\lambda+\delta\lambda)$
by a first-order canonical transformation 
\begin{align}
I(\lambda+\delta\lambda)-I(\lambda)=d_{\lambda}I(\lambda) & =\left\{ I(\lambda),G(I,\theta,\lambda)\right\} ,\nonumber \\
\phi(\lambda+\delta\lambda)-\phi(\lambda)=d_{\lambda}\phi(\lambda) & =\left\{ \phi(\lambda),G(I,\theta,\lambda)\right\} .
\end{align}
The generating function can be written as a sum of a secular term
and a completely periodic term

\begin{equation}
G(I,\theta,\lambda)=W(I,\theta,\lambda)+\alpha(I,\lambda),
\end{equation}
where $W$ is given by (\ref{W}). The Eq. (\ref{dI}) follows from 

\begin{align*}
\left\langle d_{\lambda}\phi(\lambda)\wedge d_{\lambda}I(\lambda)\right\rangle  & =\left\langle \partial_{\mu}\phi\partial_{\upsilon}I-\partial_{\upsilon}\phi\partial_{\mu}I\right\rangle d\lambda_{\mu}\wedge d\lambda_{\nu}\\
 & =\left\langle -\frac{\partial G_{\mu}}{\partial I}\frac{\partial G_{\nu}}{\partial\phi}+\frac{\partial G_{\mu}}{\partial I}\frac{\partial G_{\nu}}{\partial\phi}\right\rangle d\lambda_{\mu}\wedge d\lambda_{\nu}\\
 & =\left\langle \left\{ G_{\mu},G_{\nu}\right\} \right\rangle d\lambda_{\mu}\wedge d\lambda_{\nu}\\
 & =\left\langle \left\{ W_{\mu},W_{\nu}\right\} \right\rangle d\lambda_{\mu}\wedge d\lambda_{\nu}.\blacksquare
\end{align*}
Notice that $W$ gives a parallel transport condition among infinitesimally
close Tori

\[
\left\langle \phi_{W}(\lambda+\delta\lambda)-\phi(\lambda)\right\rangle =\left\langle \left\{ \phi,W\right\} \right\rangle =0.
\]


\begin{thebibliography}{10}
\bibitem{berry}M. V. Berry, Proc. R. Soc. London $\mathbf{A392}$,
457 (1984).

\bibitem{WZ}F. Wilczek and A. Zee. Phys. Rev. Lett. $\mathbf{52}$
(1984).

\bibitem{hannay}J. H. Hannay, J. Phys. A $\mathbf{18}$, 221 (1985).

\bibitem{hannay2}M.V. Berry, J. Phys. A $\mathbf{18}$, 15 (1985).

\bibitem{hannay3}R. Montgomery, Comm. Math. Phys. $120$ (1988).

\bibitem{berryhannay}H. D. Liu, S. L. Wu, and X. X. Y, Phys. Rev.
A $\mathbf{83}$, 062101

\bibitem{berryhannay2}Y.Liu, Y. N. Zhang, H. D. Liu, and H. Y. Sun,
Front. Phys. 10:1090973, (2023).

\bibitem{KvN1}B. O Koopman, Proc. Natl. Acad. Sci. U.S.A. $\mathbf{17}$,
315 (1931).

\bibitem{KvN2}J. von Neumann, Ann. Math. $\mathbf{33}$, 587 (1932).

\bibitem{KvN3}D. Mauro, ``Topics in Koopman\textendash von Neumann
theory'', Ph.D. Thesis Universita degli Studi di Trieste (2002), quant-ph/0301172.

\bibitem{KvN4}D. Bondar, et al, Phys. Rev. Lett. $\mathbf{109}$,
190403 (2012).

\bibitem{KvN5}A. D. Berm\'udez Manjarres, J. Phys. A: Math. Theor.
$\mathbf{55}$ 405201 (2022).

\bibitem{klein}U. Klein, Quantum Stud.: Math. Found. $\mathbf{5}$,
219 (2018).

\bibitem{arnoldergodic}V. I Arnold and A. Avez,\textquotedblleft Ergodic
problems in classical mechanics\textquotedbl{} (Benjamin, New York,
1968).

\bibitem{polavieja}G.G. de Polavieja. ``Geometric phase and angle
for noncyclic adiabatic change, revivals and measures of quantal instability''.
PhD Thesis (1999). 

\bibitem{polavieja 2}G.G. de Polavieja, Phys. Rev. Lett. $\mathbf{81}$,
1 (1998).

\bibitem{wignerKvN}R. Cabrera, D. I. Bondar, H. A. Rabitz, Phys.
Rev. A $\mathbf{88}$, 052108 (2013).

\bibitem{wignerKvN2}G. McCaul, D. V. Zhdanov, D. I. Bondar, arXiv:2302.13208.

\bibitem{adriving}A. D. Berm\'udez Manjarres and A. Botero, arXiv:2305.01125v1.

\bibitem{adriving2}M. V. Berry, J. Phys. A: Math. Theor. $\mathbf{42}$,
365303 (2009).

\bibitem{adriving3}A. del Campo Phys. Rev. Lett. $\mathbf{111}$,
100502 (2013)

\bibitem{adriving4}K. Takahashi Phys. Rev. E $\mathbf{87}$, 062117
(2013).

\bibitem{adriving5}S. Deffner, C. Jarzynski, and A. del Campo, Phys.
Rev. X $\mathbf{4}$, 021013.

\bibitem{adriving6}C. Jarzynski Phys. Rev. A $\mathbf{88}$, 040101(R)
(2013).

\bibitem{angleaction} J. Wilkie and P. Brumer, Phys. Rev. A $\mathbf{55}$,
27 (1997); Phys. Rev. A $\mathbf{55}$, 43 (1997)

\bibitem{angleaction2}T. O de Carvalho and M. A. M. de Aguiar, J.
Phys. A: Math. Gen. $\mathbf{29}$ (1996) 3597\textendash 3615

\bibitem{prigogine}I. Prigogine, ``Non-equilibrium statistical mechanics'',
p.22-30, (Dover Publications, Inc. Mineola, New York, 2017).

\bibitem{arnold}V. Arnold, ``Mathematical Methods of Classical Mechanics
2nd ed'', p. 299 (Springer-Verlag, New York, 1989)

\bibitem{continuous} M. Maamache and Y. Saadi, Phys. Rev. Lett. $\mathbf{101}$,
150407 (2008).

\bibitem{simon}B. Simon. Phys. Rev. Lett. $\mathbf{51}$ (1983).

\bibitem{messiah}W. Greiner, ``Quantum mechanics: An Introduction,
4th edition'', p. 105 (Springer-Verlag, Berlin, Heidelberg, 1989).

\bibitem{cruchinski}D. Chruscinski and A. Jamiolkowski, ``Geometric
Phases in Classical and Quantum Mechanics'', (Birkh\"{o}user, Boston,
2004).

\bibitem{ferraz}S. Ferraz-Melo, \textquotedblleft Canonical Perturbation
Theories: Degenerate Systems and Resonance\textquotedblright, (Springer-Verlag,
New York, 2007).

\bibitem{hamiltonianp}A. D. Berm\'udez Manjarres, arXiv:2107.07050v2.

\bibitem{arnold 2}V. I. Arnold, V. V. Kozlov, A. I. Neishtadt, ``Mathematical
Aspects of Classical and Celestial Mechanics 3rd. ed'', p. 175 (
Springer-Verlag, Berlin, Heidelberg, 2006).

\bibitem{robbins}J. M. Robbins, and M. V. Berry, Proc. R. Soc. A
$\mathbf{436}$, (1992) 631-661.

\bibitem{jar}C. Jarzynski, Phys. Rev. Lett. $\mathbf{74}$, 1732
(1995).
\end{thebibliography}
\end{document}